\documentclass{ws-ijmpa}

\newcommand{\DO}{\mbox{D\O}}

\begin{document}

\markboth{\DO SAR}
{Performance of an Operating High Energy Physics Data Grid: \DO SAR-Grid}

%
\catchline{}{}{}{}{}
%

\title{Performance of an Operating High Energy Physics Data Grid: \DO SAR-Grid}


\author{\footnotesize B.~Abbott,$^a$ P.~Baringer,$^b$ T.~Bolton,$^c$ 
Z.~Greenwood,$^d$ E.~Gregores,$^e$ H.~Kim,$^f$ C.~Leangsuksun,$^d$ 
D.~Meyer,$^f$ N.~Mondal,$^g$ S.~Novaes,$^e$ B.~Quinn,$^h$ 
H.~Severini\footnote{Corresponding author. Tel.: +1-405-325-3961x36415; Fax: +1-405-325-7557; E-mail: \mbox{severini@ou.edu}.},$^{,a}$
P.~Skubic,$^a$ J.~Snow,$^i$ M.~Sosebee,$^f$ J.~Yu $^f$}

\address{$^a$University of Oklahoma, Norman, OK 73019, USA\\
$^b$University of Kansas, Lawrence, KS 66045, USA\\
$^c$Kansas State University, Manhattan, KS 66506, USA\\
$^d$Louisiana Tech University, Ruston, LA 71272, USA\\
$^e$Universidade Estadual Paulista, Sao Paulo, Brazil\\
$^f$University of Texas at Arlington, Arlington, TX 76019, USA\\
$^g$Tata Institute of Fundamental Research, Bombay, India\\
$^h$The University of Mississippi, University, MS 38677, USA\\
$^i$Langston University, Langston, OK  73050, USA} 

\maketitle

\pub{Received (Day Month Year)}{Revised (Day Month Year)}

\begin{abstract}
The \DO\ experiment at Fermilab's Tevatron will record several petabytes of data over the next five years in pursuing the goals of understanding nature and searching for the origin of mass. Computing resources required to analyze these data far exceed capabilities of any one institution. Moreover, the widely scattered geographical distribution of \DO\ collaborators poses further serious difficulties for optimal use of human and computing resources. These difficulties will exacerbate in future high energy physics experiments, like the LHC. The computing grid has long been recognized as a solution to these problems. This technology is being made a more immediate reality to end users in \DO\ by developing a grid in the \DO\ Southern Analysis Region (\DO SAR), \DO SAR-Grid, using all available resources within it and a home-grown local task manager, McFarm. We will present the architecture in which the \DO SAR-Grid is implemented, the use of technology and the functionality of the grid, and the experience from operating the grid in simulation, reprocessing and data analyses for a currently running HEP experiment.

\end{abstract}

\section{Introduction}

Particle physicists employ high energy particle accelerators and complex detectors to probe ever smaller distance scales in an attempt to understand the nature and origin of the universe.  The Tevatron proton-anti proton collider located in the Department of Energy's Fermi National Accelerator Laboratory,\cite{fnal} Batavia, Illinois, currently operates at the "energy frontier" and has an opportunity of providing new discoveries through the \DO\cite{d0} and CDF\cite{cdf} high energy physics (HEP) experiments. Future experiments, such as ATLAS\cite{atlas} at the Large Hadron Collider (LHC)\cite{lhc} in the European Organization for Nuclear Research (CERN)\cite{cern} and a proposed electron-positron linear collider (LC)\cite{lc} have the prospects of either building on discoveries made at Fermilab, or making the discoveries themselves if nature has placed these new processes beyond the energy reach of the Tevatron.

\section{\DO SAR}

As part of the \DO\ experiment's effort to utilize grid technology for the expeditious analysis of data, twelve universities have formed a regional virtual organization (VO), the \DO\ Southern Analysis Region (\DO SAR).\cite{d0sar}  The centerpiece of \DO SAR is a data and resource hub called a Regional Analysis Center (RAC),\cite{d0rac} constructed at the University of Texas at Arlington with the support of NSF MRI funds.  Each \DO SAR member institution constructs an Institutional Analysis Center (IAC), which acts as a gateway to other RACs and to the grid for the users within that institution.   These IACs combine dedicated rack-mounted servers and personal desktop computers into a local physics analysis cluster. The data access system for \DO\ offline analyses is managed by a database and cataloging system called Sequential Data Access via Metadata, or SAM.\cite{sam} The MC\_Runjob package\cite{mcrunjob} provides a low-level MC manager that coordinates data files used by and produced with the executables of the MC chain, via scripts produced for the job. This package is combined with MC task management software called McFarm\cite{mcfarm}, developed at UTA, for automated, highly efficient MC production.

\section{\DO SAR-Grid}

In order to pursue full utilization of grid concepts, we are establishing an operational regional grid called \DO SAR-Grid using all available resources, including personal desktop computers and large dedicated computer clusters. We have constructed and are operating \DO SAR-Grid utilizing a framework called SAM-Grid\cite{samgrid} being developed at Fermilab.  \DO SAR-Grid will subsequently be made interoperable under other grid frameworks such as LCG,\cite{lcg} TeraGrid,\cite{teragrid} Grid3,\cite{grid3} and Open Science Grid.\cite{osg}  Wherever possible, we will exploit existing software and technological advances made by these other grid projects. We plan to develop and implement tools to support easy and efficient user access to the grid and to ensure its robustness. Tools to transfer binary executables and libraries efficiently across the grid for environment-independent execution will be developed. \DO SAR will implement the Grid for critical physics data analyses, while at the same time subjecting grid computing concepts to a stress test to its true conceptual level, down to personal desktop computers. Many of the proponents of this project are members of the LHC ATLAS\cite{atlas} experiment and a future experiment under study by the American Linear Collider Physics Group,\cite{alcpg} and will apply the experience gained from \DO SAR to these projects.

\section{Conclusions}

Experimental particle physics has always pushed information technology in directions that have had profound broader impact, with the most spectacular example being CERN's development of the World Wide Web, originally intended to facilitate communication among collaborating physicists spread around the globe. \DO SAR's goal of efficiently using computing resources scattered across the south-central United States for tasks such as complex data reduction and simulation will produce grid computing solutions with broad applicability.  This project will also likely affect how we disseminate and analyze data from future experiments at facilities such as the LHC. \DO SAR will help improve the cyber-infrastructure within the region; five participating institutions are located in states that are traditionally under-funded for R\&D activities. 
In addition to technology development, there will also be significant development of human resources in the region, delivering interdisciplinary training in physics and computer science. Information for the broader public will be disseminated on the web.
 Finally, \DO SAR will realize the conceptual grid to the level of personal desktop computers, demonstrating the performance of a grid at its fullest level.  The \DO SAR Grid will be exploited in critical physics analyses of real and simulated data from the \DO\ experiment, and later from LHC and other future experiments. These data will serve to advance our understanding of nature at the smallest distance scales.

\appendix

\end{document}